\documentclass[aps,pra,twocolumn,showpacs,amsmath,amssymb,floatfix,footinbib,superscriptaddress]{revtex4-1}
\usepackage{amsmath,natbib,graphicx,subfigure,amssymb,graphics,amsmath,mathrsfs,CJK,color}
\usepackage{multirow,fancyhdr,color,bm,tabularx,psfrag,dcolumn}
\usepackage[colorlinks=true,linkcolor=blue,urlcolor=blue,citecolor=blue]{hyperref}
\usepackage{tikz,xcolor,hyperref}
\definecolor{lime}{HTML}{A6CE39}
\DeclareRobustCommand{\orcidicon}{%
    \begin{tikzpicture}
    \draw[lime, fill=lime] (0,0)
    circle [radius=0.16]
    node[white] {{\fontfamily{qag}\selectfont \tiny ID}};\draw[white, fill=white] (-0.0625,0.095)
    circle [radius=0.007];
    \end{tikzpicture}
    \hspace{-2mm}}
\foreach \x in {A, ..., Z}
{\expandafter\xdef\csname orcid\x\endcsname{\noexpand\href{https://orcid.org/\csname orcidauthor\x\endcsname}{\noexpand\orcidicon}}}

\usepackage[left]{lineno}
\begin{document}

\title{Manipulative Properties of the Asymmetry Double Quantum Dots via Laser and Gate Voltage }

\author{Shun-Cai Zhao\orcidA{}}
\email{zscnum1@126.com.}
\affiliation{Engineering Research Center for Nanotechnology, Nanchang University, Nanchang 330047, PR China}
\affiliation{Institute of Modern Physics,Nanchang University,Nanchang 330031, PR China}
\affiliation{School of Materials Science and Engineering, Nanchang University, Nanchang 330031, PR China}

\author{Zheng-Dong Liu}
\email{lzdgroup@ncu.edu.cn.}
\affiliation{Engineering Research Center for Nanotechnology, Nanchang University, Nanchang 330047, PR China}
\affiliation{Institute of Modern Physics,Nanchang University,Nanchang 330031, PR China}
\affiliation{School of Materials Science and Engineering, Nanchang University, Nanchang 330031, PR China}
 
\begin{abstract}
We present the density matrix approach for theoretical description
of an asymmetric double quantum dot(QD)system.The results show that
the properties of gain,absorption and dispersion of the double QD
system ,the population of the state with one hole in one dot and an
electron in other dot transferred by tunneling can be manipulated by
the laser pulse or gate voltage.Our scheme may propose a probability
of electro-optical manipulation of quantum system.
\begin{description}
\item[PACs]{ 78.67.Hc ; 42.50.Gy ; 73.21.La}
\item[Keywords]{Quantum dot; Tunneling; Absorption and dispersion; Average occupation.}
\end{description}
\end{abstract}
\maketitle

\section{Introduction}
Quantum dot(QD)structures provide a three-dimensional confinement of
carriers.Electrons and holes in the QD can occupy only a set of
states with discrete energies. For this reason the physics of
quantum dots show many parallels with the behavior of atoms, and
quantum dots are often referred to as artificial atoms. Thus,the QD
can be used to perform"atomic physics"experiments in solid-state
structures$^{[1]}$. One advantage of QDs is that they provide
different energy scales and physical features which can be easily
varied over a wide range of values. Most importantly,it's that QDs
allow the control of direct quantum-mechanical electronic coupling
with not only composition but external voltages$^{[2,3]}$ .The
flexible systems indicate the ideal for theoretical and experimental
investigations,where the interactions between light and matter can
be studied in a fully controlled and well characterized environment.
These features make semiconductor QD a lot of possible applications
in electro-optical devices $^{[4,5]}$ and quantum-information
processing$^{[6\sim11]}$. In the latter case, one can exploit the
optical excitation in a QD$^{[7,12\sim13]}$ or its spin
state$^{[6,8]}$ as qubits . These expectations are produced by
experimental advances in the coherent manipulation of QD states,
such as the exciton Rabi oscillations in single dots achieved by the
application of electromagnetic pulses$^{[9,14\sim17]}$. Coherent
phenomena in ensembles of QDs have also been
observed$^{[18\sim23]}$.

In the present study, we investigate the manipulative properties of
an asymmetric double QD . The system consists of two dots with
different band structures coupled by tunneling,and it can be
fabricated using self-assembled dot growth technology$^{[24]}$. With
the application of an electromagnetic field an electron is excited
from the valence band to the conduction band of one of the quantum
dots.The electron can be transferred by tunneling to the other
quantum dot. Applied a gate voltage,the conduction-band levels get
closer to resonance and their coupling increases.The effective
decoupling of the valence-band levels will occur when those levels
get more off-resonance. In this paper,The density matrix approach is
used for the theoretical description of the double QD system .We
analyze the system's absorption and dispersion properties ,the
occupation of the state with the electron-hole pair distributing in
two different dots manipulated by external laser pulse or gate
voltage.

\section{Theoretical model}

\begin{figure}[h!]
\center{
\includegraphics[width=3in]{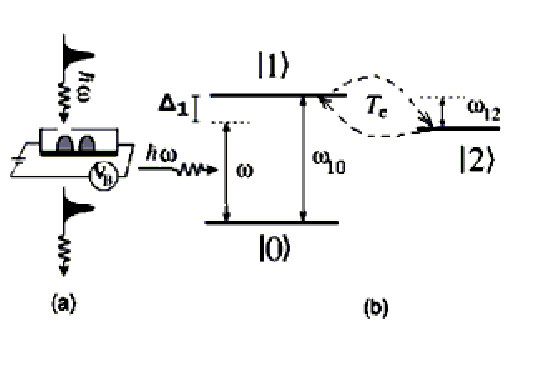}
\caption{(a)Schematic of the setup.An optical pulse transmits
the left QD.$V_{B}$ is a gate voltage.(b) Schematic band structure
and level configuration for the double QD system.A pulsed laser
excites one electron from the valence band that can tunnel to the
other dot.$|0\rangle$is the system without excitations,$|1\rangle$a
pair of electron and hole bound in the first dot,and$|2\rangle$one
hole in the first dot with an electron in the second dot.\label{f1}}}
\end{figure}

Fig.1 depicts a diagram of the asymmetrical double QD for our
model$^{[25]}$. The asymmetrical double QD consists of two dots (the
left one and the right one in Fig.1(a)) with different band
structures coupled by tunneling. The lateral geometry enables
coupling in two dimensions, and thus allows for a natural upscaling
to very large numbers of quantum gates across a semiconductor
substrate. Lateral quantum coupling between two distinct
self-assembled quantum dots has been observed$^{[26]}$, and it
demonstrates that the QD molecules consisting two distinct dots,
even though the QDS are separated by only a few nanometers of
barrier material. At nanoscale interdot separations, the exciton
states consist of delocalized electron states,i.e.,holes are
localized in the QDs,while electrons become almost entirely
delocalized in the QD molecules.In Fig.1(b),the energy-level diagram
of an asymmetric double QD is depicted by three states: the ground
state$|0\rangle$ , where the system has no excitations, the exciton
state$|1\rangle$, where a pair of an electron and a hole are bound
in the first quantum dot$^{[14\sim17,27]}$, and finally the indirect
exciton state $|2\rangle$, where the hole is in the first quantum
dot and the electron is in the second quantum dot.

The electromagnetic field is introduced by dipole interaction,which
couples the ground state $|0\rangle$ with the exciton state
$|1\rangle$. With the application of an electromagnetic field an
electron is excited from the valence band to the conduction band in
the left dot, which can also tunnel to the right dot.The electron
tunneling couples the exciton state $|1\rangle$ with the indirect
exciton state $|2\rangle$. Placing a gate electrode between the two
quantum dots one can control the tunnel barrier in this
structure$^{[25]}$.Without the gate voltage, the conduction-band
electron energy levels are out of resonance and the electron
tunneling between two quantum dots is very weak.While the gate
voltage is present,the conduction-band electron levels come close to
resonance and the electron tunneling between the two quantum dots is
significantly enhanced$^{[8]}$. The valence-band energy levels
become more off-resonant and the hole tunneling can be neglected.
Using this configuration the dynamics of the system is described by
the following density matrix equations$^{[28]}$,

\begin{eqnarray}
\dot{\rho_{01}}=i(\triangle_{1}+i\gamma_{1})\rho_{01}-i\frac{\Omega}{2}(\rho_{11}-\rho_{00})+iT_{e}\rho_{02}\\
\dot{\rho_{12}}=-i(\triangle_{1}-\triangle_{2}-i\gamma_{2})\rho_{12}-i\frac{\Omega}{2}\rho_{02}-iT_{e}(\rho_{22}-\rho_{11})\\
\dot{\rho_{02}}=i(\triangle_{2}+i\gamma_{3})\rho_{02}-i\frac{\Omega}{2}\rho_{12}+iT_{e}\rho_{02}\\
\dot{\rho_{00}}=\Gamma_{20}\rho_{22}+\Gamma_{10}\rho_{11}-i\frac{\Omega}{2}(\rho_{10}-\rho_{01})\\
\dot{\rho_{11}}=-(\Gamma_{10}+\Gamma_{12})\rho_{11}+i\frac{\Omega}{2}(\rho_{10}-\rho_{01})-iT_{e}(\rho_{21}-\rho_{12})
\end{eqnarray}

with $\rho_{00}+\rho_{11}+\rho_{22}=1$ , $\rho_{ij}$=
$\rho_{ji}^{\ast }$ , and with $i\neq j$, i , j = 0, 1, 2 . Here
$\Omega$ = $-\vec{\mu_{01}}\hat{\varepsilon}\textbf{E}/\hbar$  is
the Rabi frequency of the laser pulse(with the angular frequency
$\omega$) to drive the transition $|0\rangle\leftrightarrow
|1\rangle$, where $\vec{\mu}_{01}$ is the associated dipole
transition matrix element. And  $\hat{\varepsilon}$ is the
polarization vector and E is the electric field amplitude of the
laser pulse .$\Delta_{1}=\omega_{10}-\omega$ denotes the detuning of
the electromagnetic field of the laser pulse from resonance with the
transition $|0\rangle\leftrightarrow|1\rangle$ .The detuning
$\Delta_{2}$ is defined as $\Delta_{2}=\Delta_{1}-\omega_{12}$,where
$\omega_{ij}=\omega_{i}-\omega_{j}$,with the energies of the
$|i\rangle$ and $|j\rangle$ states being $\hbar\omega_{i}$ and
$\hbar\omega_{j}$. $T_{e}$ is the electron tunneling coupling
coefficient.The parameters $T_{e}$  and $\omega_{12}$ can be simply
tuned with gate voltage. Finally, $\Gamma_{ij}$ denotes the decay
rate of the populations from state $|i\rangle$ to state $|j\rangle$
, and  $\gamma_{1},\gamma_{2},\gamma_{3}$ depict the decay rates of
coherence of the off-diagonal density matrix element for
$\rho_{10},\rho_{12}$ and $\rho_{20}$, respectively.

\section{Results and discussion}

\begin{figure}[h!]
  \centering
  \includegraphics[width=2.5in]{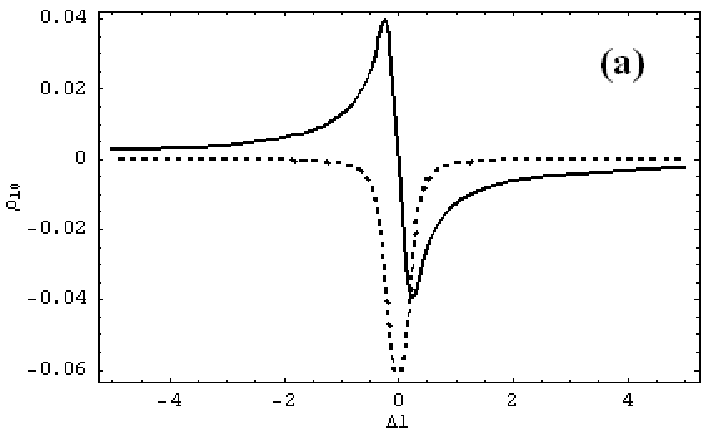}%
  \hspace{0in}%
  \includegraphics[width=2.5in]{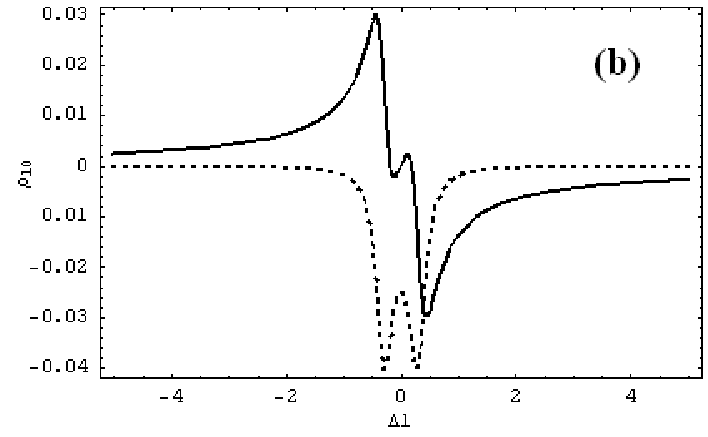}
  \hspace{0in}%
  \includegraphics[width=2.5in]{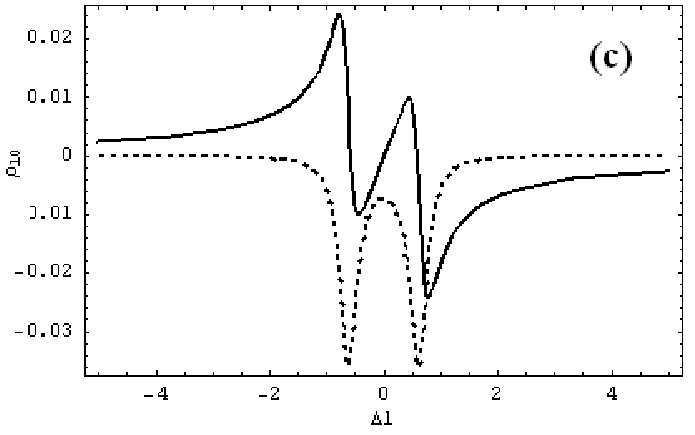}
  \hspace{0in}%
 \includegraphics[width=2.5in]{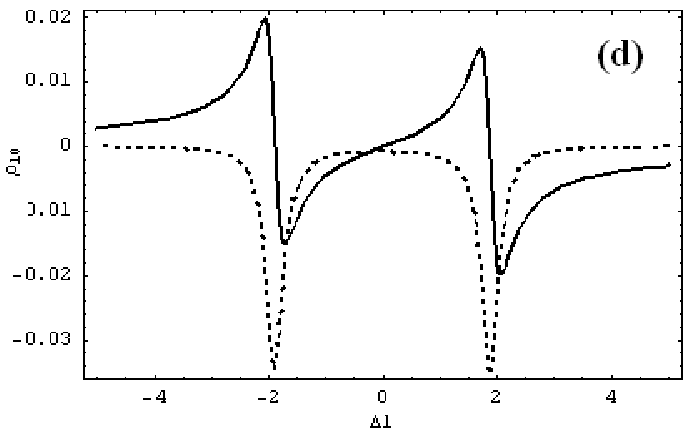}
 \hspace{0in}%
 \includegraphics[width=2.5in]{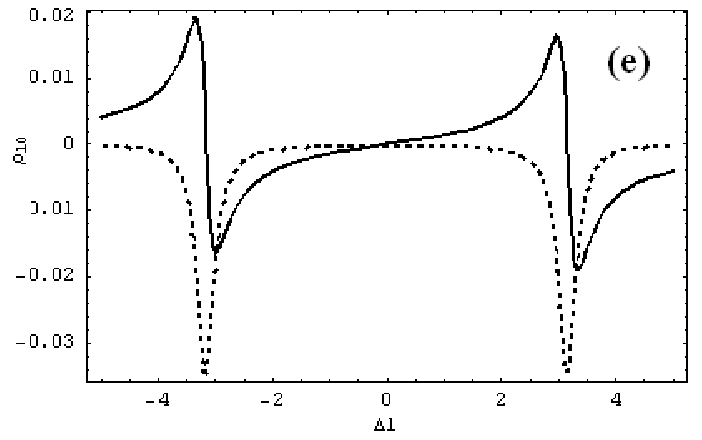}
\caption{Real and imaginary parts of $\rho_{10}$ as a function of the
detuning $\Delta_{1}$.The dashed and solid curves are the imaginary
and real parts of $\rho_{10}$,respectively.The values of $T_{e}$ in
(a)$\rightarrow$(e) are 0.5$\gamma$,1$\gamma$,2$\gamma$,6$\gamma$,10$\gamma$. \label{f2}}
\end{figure}

We now discuss the absorption and dispersion properties of the
system by numerical solution of Eqs.(1-5) under the steady-state
condition. For this purpose we will examine the coherence term
$\rho_{10}$ for the laser pulse in terms of its real and imaginary
parts as a function of $\Delta_{1}$. The imaginary (real) part of
$\rho_{10}$ versus $\Delta_{1}$ represents the double QD system's
absorption (dispersion) spectrum to the laser pulse. In Fig.2,we
concentrate on the situation of exact resonance between states
$|1\rangle$ and $|2\rangle$,i.e.,$\omega_{12}=0$.Other parameters
are scaled by $\gamma$:$\gamma_{1}= \gamma_{2}=\gamma,\gamma_{3}=
0.25 \gamma$,$\Gamma_{10}=\Gamma_{12 }=\Gamma_{20}=\Omega=
0.5\gamma$,$\Delta_{2}=\Delta_{1}\gamma$.Typical values of for
realistic quantum dot structures can be found in Refs$^{[29-31]}$.
The plots of Im($\rho_{10}$) and Re($\rho_{10}$) as a function of
$\Delta_{1}$ are shown with $T_{e}$=0.5$\gamma$,
1$\gamma$,2$\gamma$,6$\gamma$,10$\gamma$ in Fig.2(a)
$\leftrightarrow$(e) to investigate the the absorption and
dispersion characteristic dependence of the electron tunneling
coupling coefficient $T_{e}$.Which can be simply tuned with external
gate voltage.

The solid and dashed curves correspond to the real and imaginary
parts of $\rho_{10}$ in Fig.2.The profiles of dash curve show
absolute gain versus $\Delta_{1}$ in Fig.2(a)$\leftrightarrow$(e),
when the electron tunneling coupling coefficient $T_{e}$ is tuned
with a set of values of the external gate voltage. In Fig.2(a),we
can see a gain vale appears at the resonant
point$(\Delta_{1}=0)$,when $T_{e}$=0.5 $\gamma$.The value of the
electron tunneling coupling coefficient $T_{e}$ increases to double
in Fig.2(b),the gain vale at the resonant point splits into two .
The degree of gain at the resonant location decreases continually in
Fig.2(c).The two gain vales move to two symmetric points
$\Delta_{1}=\pm2$ and a transparency window occurs the resonant
location when $T_{e}$=6 $\gamma$. The transparency window widens
largely,the locations of the two gain vales are about
$\Delta_{1}=\pm3.25$ in Fig.2(e).In other words,increasing of the
values of the external gate voltage would inhibit the gain at the
resonant location. This may be a scheme of the voltage control of
optics to realize transparency. The slope of the solid curve
displays the dispersion property of the double QD system. The value
of the slope of the solid curve is negative  and large at the
resonant location in Fig.2(a). It means an intensive anomalous
dispersion at at the resonant position. In Fig.2(b)
$\leftrightarrow$(e)the slope changes into positive.And the values
decrease with the variation of  $T_{e}$= 1$\gamma$,2$\gamma$,
6$\gamma$,10$\gamma$. It shows that the intensity of normal
dispersion weakens gradually.varying the gate voltage can effect the
absorption and dispersion properties simultaneously .

\begin{figure}[h!]
\center{
\includegraphics[width=3.5in]{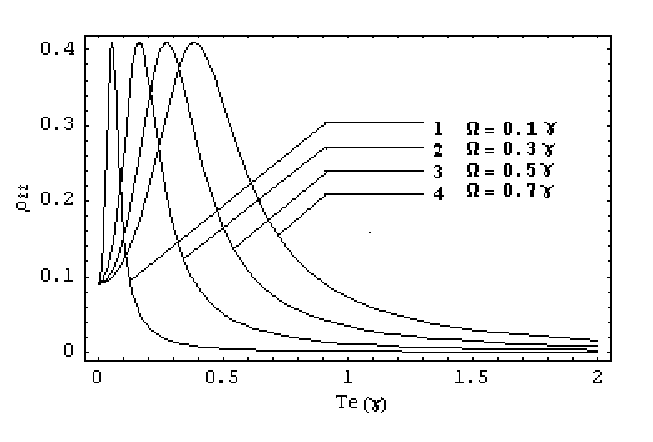}
\caption{The occupation of state$|2\rangle$as a function of the electron
tunneling coupling coefficient $T_{e}$ with a set of the intensity
of the laser pulse  $\Omega$  being given. \label{f3}}}
\end{figure}

Another interesting phenomenon is the photocurrent. The photocurrent
signal induced by the laser pulse in the double QD would be a direct
measure of how much the electron has tunneled to the second dot.
Experimentally this could be very useful since one could monitor the
population of the second dot as a nondisturbing probe of the
coherent state of the QD. Fig.3 and Fig.4 give the population of the
state state $|2\rangle$ with the parameter values
$\omega_{10}=0.2\gamma,\Delta_{1}=0$.Other parameters are the same
as those in Fig.2.

Keeping the intensity of the laser pulse($\Omega$)fixed and varying
the gate voltage are possible in experiment. The electron tunneling
coupling coefficient $T_{e}$ is then tuned with gate voltage.Fig.3
depicts the case.We can see the population of the state $|2\rangle$
is not a monotonic function of the electron tunneling coupling
coefficient $T_{e}$ .Remaining the Rabi frequency $\Omega$=0.1
$\gamma$ ,we notice that the population of the state $|2\rangle$ is
about 0.1 when $T_{e}$=0.This demonstrates that interdot tunneling
can transfer some of the population of the indirect exciton state
$|2\rangle$ under the condition of without gate voltage.When the
electron tunneling coupling coefficient $T_{e}$ increases to
0.05$\gamma$,the population of the state $|2\rangle$ comes to a
sharp peak.The peak reflects a maximum transfer of population to the
indirect exciton state $|2\rangle$.If we further increase $T_{e}$,we
observe a suppression of the population,and the value of the
population basically tends to zero in the end.Varying the values of
the Rabi frequency with $\Omega$=0.3$\gamma$,
0.5$\gamma$,0.7$\gamma$,we can see the profiles of the population of
the state $|2\rangle$ are alike.And the peak values all are 0.4. But
the locations of the peak are different with the variation of the
Rabi frequency. In other words, the intensity of the laser pulse
can't alter the peak of the population of the state $|2\rangle$ but
can change the location of the peak.

\begin{figure}[h!]
\center{
\includegraphics[width=3.5in]{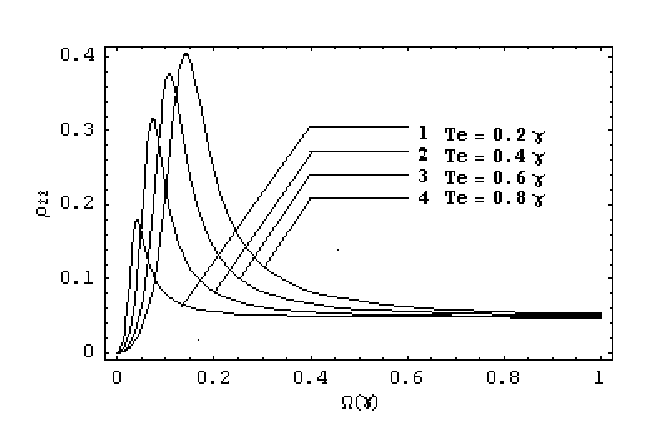}
\caption{The occupation of
state$|2\rangle$as a function of intensity of the laser pulse
$\Omega$ with a set of the electron tunneling coupling coefficient
Te being given. \label{f4}}}
\end{figure}

The intensity of the laser pulse is another tunable parameter
experimentally. It's possible to keep the gate voltage fixed and
vary the intensity of the laser pulse. Fig.4 shows that the
population of the state $|2\rangle$ with the given set of  $T_{e}$
values is not a monotonic function of $\Omega$.It's interesting to
note that when the Rabi frequency $\Omega$=0,the population of the
state $|2\rangle$ is zero.Which is contrarily different with
Fig.3.This reflects zero transfer of population to the indirect
exciton state $|2\rangle$. When the electron tunneling coupling
coefficient $T_{e}$=0.2 $\gamma$,there is a peak at the point where
$\Omega$$\approx$ $0.05\gamma$ ,and the value of the peak is 0.18.
Further increasing the the Rabi frequency,we observe a suppression
of the population of the state $|2\rangle$, where the value drops to
0.06,approximately.From the profiles of Fig.4,we can see the values
of the peak are 0.32,0.38,0.4 with the electron tunneling coupling
coefficient $T_{e}$=0.4$\gamma$,0.6$\gamma$,0.8$\gamma$.Which is
also different with Fig.3.The increasing of the value of the peak
displays more transfer of population to the indirect exciton state
$|2\rangle$. The results show that the photocurrent is controllable
with a direct measure of how much the electron has tunneled to the
second dot. The electron population of the state $|2\rangle$ can be
enhanced firstly , suppressed later. This  provides an
electro-optical method to control the electron population of the
second dot.These propose an electro-optical controllable method to
control the electron population of the second dot.

\section{Conclusion}

We have theoretically investigated the gain,absorption and
dispersion properties,the population of the indirect exciton state
$|2\rangle$ of the asymmetric double QD system. Increasing the
electron tunneling coupling coefficient $T_{e}$, the gain vale
decreases to a transparency window at the resonant location and the
variation of from anomalous to normal dispersion occurs.And the
intensive of normal dispersion decreases gradually.The value of the
population of state $|2\rangle$ can be manipulated by the laser
pulse intensity and gate voltage.The population of state $|2\rangle$
can be enhanced firstly,then suppressed later. However, Fixing the
intensity of the laser pulse and varying the gate voltage ,the
population is nonzero when $T_{e}$ =0.And the population basically
tends to zero if we further increase $T_{e}$. Keeping the gate
voltage fixed and varying the intensity of the laser pulse,the
population is zero when $\Omega$=0.The population basically tends to
a constant if we further increase $\Omega$. These may propose an
electro-optical controllable method to control the quantum optics
system. And this manipulation could have a profound impact in the
field of quantum information processing.

\section*{Acknowledgments}

This work has been supported by the National Natural Science
Foundation of China (Grant No.60768001 and No.10464002).


\begin{thebibliography}{0}%
\makeatletter
\providecommand \@ifxundefined [1]{%
 \@ifx{#1\undefined}
}%
\providecommand \@ifnum [1]{%
 \ifnum #1\expandafter \@firstoftwo
 \else \expandafter \@secondoftwo
 \fi
}%
\providecommand \@ifx [1]{%
 \ifx #1\expandafter \@firstoftwo
 \else \expandafter \@secondoftwo
 \fi
}%
\providecommand \natexlab [1]{#1}%
\providecommand \enquote  [1]{``#1''}%
\providecommand \bibnamefont  [1]{#1}%
\providecommand \bibfnamefont [1]{#1}%
\providecommand \citenamefont [1]{#1}%
\providecommand \href@noop [0]{\@secondoftwo}%
\providecommand \href [0]{\begingroup \@sanitize@url \@href}%
\providecommand \@href[1]{\@@startlink{#1}\@@href}%
\providecommand \@@href[1]{\endgroup#1\@@endlink}%
\providecommand \@sanitize@url [0]{\catcode `\\12\catcode `\$12\catcode
  `\&12\catcode `\#12\catcode `\^12\catcode `\_12\catcode `\%12\relax}%
\providecommand \@@startlink[1]{}%
\providecommand \@@endlink[0]{}%
\providecommand \url  [0]{\begingroup\@sanitize@url \@url }%
\providecommand \@url [1]{\endgroup\@href {#1}{\urlprefix }}%
\providecommand \urlprefix  [0]{URL }%
\providecommand \Eprint [0]{\href }%
\providecommand \doibase [0]{http://dx.doi.org/}%
\providecommand \selectlanguage [0]{\@gobble}%
\providecommand \bibinfo  [0]{\@secondoftwo}%
\providecommand \bibfield  [0]{\@secondoftwo}%
\providecommand \translation [1]{[#1]}%
\providecommand \BibitemOpen [0]{}%
\providecommand \bibitemStop [0]{}%
\providecommand \bibitemNoStop [0]{.\EOS\space}%
\providecommand \EOS [0]{\spacefactor3000\relax}%
\providecommand \BibitemShut  [1]{\csname bibitem#1\endcsname}%
\let\auto@bib@innerbib\@empty
\end{thebibliography}%


\begin{thebibliography}{00}

\bibitem{1}Empedocles  S  A ,DNorris  J  and  Bawendi  M   G  1996 {\it Phys. Rev. Lett.} {\bf 77} 3873.

\bibitem{2}Hu  Y  Z ,Koch  S  W ,Lindberg  M , Peyghambaruan  N, Pollock  R and Abraham  F  F 1990 {\it Phys. Rev. Lett.}{\bf 64} 1805.

\bibitem{3}Ohno  Y , Asaoka  K ,Kishimoto  S , Maezawa  K  and  Mizutani  T  2000 {\it J. Appl. Phys.} {\bf 87} 4333.

\bibitem{4}Shchekin  O  B , Park  G , Huffaker  D  L   and  Deppe  D  G  2000 {\it Appl.Phys. Lett.} {\bf 77} 466.

\bibitem{5}Saito  H , Nishi  K   and  Sugou  S  2001  {\it  Appl. Phys. Lett.} {\bf 78} 267.

\bibitem{6}Loss  D and  DiVincenzo  D  P  1998 {\it Phys. Rev. A } {\bf 57} 120.

\bibitem{7}Sherwin  M  S , Imamo$\check{g}$ lu  A  and Montroy  T 1999 {\it Phys. Rev. A } {\bf 60} 3508.

\bibitem{8}Awschalom  D D ,Burkard  G ,DiVincenzo  D P ,Loss  D , Sherwin  M  and  Small  A  1999  {\it Phys. Rev. Lett.} {\bf 83} 4204.

\bibitem{9}Stievater  T  H , Li  X , Steel  D  G ,Gammon  D ,Katzer  D  S ,Park  D ,Piermarocchi  C  and Sham  L  J  2001  {\it  Appl. Phys. Lett.} {\bf 87} 133603.

\bibitem{10}Nielsen  M  A ,Chuang  I  L , {\it Quantum Computation and Quantum Information} {\bf (Cambrige, England,Cambrige University Pres 2000)}.

\bibitem{11}DENG  H  L  and  Fang  X  M  2007  {\it Chin.Phys.Lett.}{\bf 24}(11) 3051.

\bibitem{12}Ekert  A  and  Jozsa  R  1996  {\it Rev. Mod. Phys.} {\bf 68} 733.

\bibitem{13}WANG  Z  B  ,ZHANG  J  Y  ,et. al. 2008 {\it Chin.Phys.Lett.}{\bf 25}(12) 4435.

\bibitem{14}Kamada  H , Gotoh  H , Temmyo  J ,Takagahara  T and  Ando  H  2001 {\it Phys. Rev. Lett.} {\bf 87} 246401.

\bibitem{15}Htoon  H , Takagahara  T ,Kulik  D , Baklenov  O , Holmes  A  L ,et. al. 2002 {\it Phys. Rev. Lett.} {\bf 88} 087401.

\bibitem{16}Zrenner  A ,Beham  E , Stufler S ,Findeis  F ,Bichler  M  and  Abstreiter  G  2002  {\it Nature (London)} {\bf 418} 612.

\bibitem{17}Li X , Wu Y , Steel  D  G , Gammon  D  and  Stievater  T  H , et.al. 2003 {\it Science} {\bf 308} 809.

\bibitem{18}Kim  J  , Benson  O , Kan  H  and  Yamamoto  Y  1999  {\it Nature(London)} {\bf 397} 500.

\bibitem{19}Michler  P , Kiraz  A  , Becher  C  ,et.al. 2000 {\it  Science} {\bf 290} 2282.

\bibitem{20}Cole  B  E  , Williams  J  B ,King  B  T ,Sherwin  M  S  and  Stanley  C  R 2001 {\it Nature(London)} {\bf 410} 60.

\bibitem{21}Borri  P , Langbein  W , Schneider  S , Woggon  U , Sellin  R  L , Ouyang  D  and  Bimberg  D  2002 {\it Phys. Rev. B} {\bf 66} 081306.

\bibitem{22}Pelton  M , Santori  C , Vu$\check{c}$kovi$\acute{v}$  J , Zhang  B , Solomon  G  S  , Plant  J  and Yamamoto  Y  2002 {\it Phys. Rev. Lett} {\bf 89} 233602.

\bibitem{23}LI  Q , ZHONG  X  W  and  HU  X  M  2008 {\it Chin.Phys.Lett.}{\bf 25}(9) 3234.

\bibitem{24}Petroff  P  M , Lorke  A  and  Imamo$\check{g}$  lu  A  2001 {\it Phys. Today} {\bf 54(5)} 46.

\bibitem{25}Villas-B$\hat{o}$s  J  M , Govorov  A  O , et.al. 2004 {\it Phys. Rev. B} {\bf 69} 125342.

\bibitem{26}Beirne  G  J  ,Hermannst$\ddot{a}$dter  C , Wang  L ,Rastelli  A  , Schmidt  O  G  and  Michler  P  2006  {\it  Phys. Rev. Lett. } {\bf 96} 137401.

\bibitem{27}Stievater  T  H , Li  X ,Steel  D  G , Gammon  D , Katzer  D  S , Park  D  ,  Piermarocchi  C  and  Sham  L  J  2001  {\it Phys. Rev. Lett.} {\bf 87} 133603.

\bibitem{28}Ioannou  M , Boviatsis  J  and  Paspalakis  E  2008  { \it Phys.E} {\bf 40} 2010.

\bibitem{29}Chang-Hasnain  C  J , Ku  P  C , Kim  J  and  Chuang  S  L  2003 {\it Proc. IEEE} {\bf 91} 1884.

\bibitem{30}Kim  J , Chuang  S  L , Ku  P  C  and  Chang-Hasnain  C  J  2004  {\it J. Phys.Condens. Matter} {\bf 16} S3727.

\bibitem{31}Kaer Nielsen  P , Thyrrestrup  H , Mork  J  and  Tromborg  B  2007  {\it Opt.Express} {\bf 15} 6396 .

\end{thebibliography}
\end{document}